\documentclass[aps,pre,twocolumn,superscriptaddress,showpacs,amsmath,amssymb,floatfix]{revtex4-2}

\usepackage{graphicx}
\usepackage{amssymb} \usepackage{amsmath} \usepackage{bm}
\usepackage[colorlinks,linkcolor=blue,citecolor=blue]{hyperref}  \usepackage{url}
\usepackage[dvipsnames]{xcolor}

\newcommand{\Pe}{{\rm Pe}}

\newcommand{\avgt}{{\langle t \rangle}}

\begin{document} 

\title{Optimal navigation of microswimmers in complex and noisy environments} 

\author{Lorenzo Piro}
\affiliation{Max Planck Institute for Dynamics and Self-Organization, 37077 G\"ottingen, Germany}

\author{Beno\^{i}t Mahault}
\affiliation{Max Planck Institute for Dynamics and Self-Organization, 37077 G\"ottingen, Germany}

\author{Ramin Golestanian}
\email{ramin.golestanian@ds.mpg.de}
\affiliation{Max Planck Institute for Dynamics and Self-Organization, 37077 G\"ottingen, Germany}
\affiliation{Rudolf Peierls Centre for Theoretical Physics, University of Oxford, Oxford OX1 3PU, United Kingdom}

\begin{abstract}
We design new navigation strategies for travel time optimization of microscopic self-propelled particles in complex and noisy environments. In contrast to strategies relying on the results of optimal control theory, these protocols allow for semi-autonomous navigation as they do not require control over the microswimmer motion via external feedback loops. Although the strategies we propose rely on simple principles, they show arrival time statistics strikingly similar to those obtained from stochastic optimal control theory, as well as performances that are robust to environmental changes and strong fluctuations. These features, as well as their applicability to more general optimization problems, make these strategies promising candidates for the realization of optimized semi-autonomous navigation. 
\end{abstract}

\maketitle 


\section{Introduction}
The problem of finding the most efficient route towards a desired target has important technological and medical applications, such targeted drug delivery at the microscale~\cite{park,manzari}, environmental monitoring~\cite{trincavelli}, and optimization of plane routes~\cite{Szczerba2000IEEE}. With the advent of theoretical and experimental prescriptions for how to make microswimmers \cite{NG:2004,Dreyfus2005Nature,Golestanian2007,Illien2017}, the fictional idea of making microscopic devices capable to delivering molecular cargo at the desired location in human body has moved closer to reality.

A solution of this classical problem can be traced back to the work of E. Zermelo~\cite{zermelo}, who derived the steering policy which minimizes the travel time of a vessel moving at constant speed in presence of wind. In the context of micoswimmers, Zermelo's work has been extended to more general contexts, including fuel consumption and dissipation minimization~\cite{liebchen}, time-varying flows~\cite{Techy2011ISR}, motion on non-Euclidean spaces~\cite{piro}, or the role of hydrodynamic interactions~\cite{daddi}. The approach of Zermelo, however, does not account for thermal fluctuations, which play a prominent role at the micro-scale. 

Optimal navigation in the presence of noise falls into the class of problems addressed by stochastic optimal control theory~\cite{pontryagin,bellman1954,yong1999book,kappen2005,pinti}. Considering a cost function defined on the configuration space ({\it e.g.} the mean arrival time to a given target starting from a specific position), a stochastic optimization principle can be used to show that it obeys a so-called Hamilton-Jacobi-Bellman equation, from which the optimal control map can be obtained~\cite{yong1999book,bertsekas}. In parallel, machine learning algorithms such as reinforcement learning provide convenient and increasingly popular routes to determining the optimal control landscapes~\cite{colabrese,Schneider2019EPL,biferale,yang2020AIS,mehlig2020,muinos2021ScienceR}. In practice, implementations of such optimal policies can then be achieved via external feedback loops for the actuation of a microswimmer motion~\cite{Tierno2008b,Bregulla2014ACS,mano,Das2015,Gomez2017scienrep,chen2018small,Fernandez2020natcomm}.

On the other hand, a number of natural and artificial microswimmers exhibit tactic behaviour~\cite{tsang2020review}, {\it i.e.} are able to adapt their motility in response to external stimuli such as light~\cite{Bennett2015,Lozano2016natcomm,Dai016NatNano}, chemical concentration~\cite{Hong2007PRL,Lagzi2010maze,saha2014PRE,Kranz:2016} 
or viscosity~\cite{Palacci2006sciadv,Kantsler2014elife} gradients, as well as magnetic~\cite{Klumpp2016magnbact,Dreyfus2005Nature,Matsunaga:2017,Meng:2018} or gravitational~\cite{Hagen2014natcomm} fields, in an autonomous fashion. Harnessing guidance provided by taxis allows microswimmers to perform complex tasks~\cite{Lagzi2010maze,Zheng2017Natcomm,Frangipane2018elife,huang2019}
in a semi-autonomous way as it does not rely on real time external feedback mechanisms.

Here, we show how these ideas can be applied to the problem of optimal navigation in complex and noisy environments. Considering a minimal but non-trivial optimization problem in two dimensions, we propose novel navigation policies inspired by the control maps provided by the exact global optimization problem, which can be implemented in a semi-autonomous fashion. Using extensive Brownian dynamics simulations, we show that the new policies show performances comparable to that obtained from stochastic optimal control theory, and demonstrate their robustness upon changes in the environment, as well as to positional and rotational fluctuations. Lastly, we illustrate how the semi-autonomous policies can be conveniently adapted to a broader class of problems such as navigation on curved manifolds.\\


\begin{figure*}[t!]
\includegraphics[width=\linewidth]{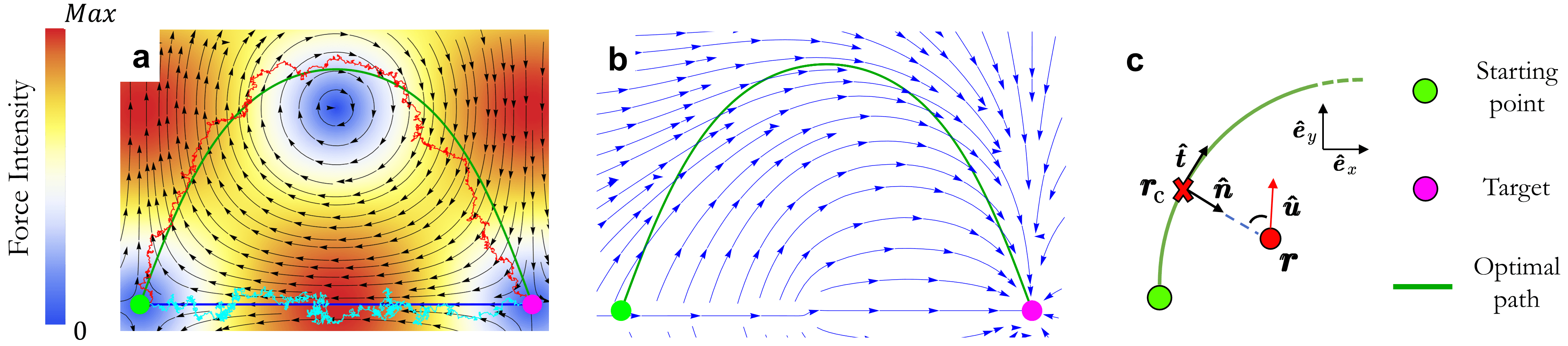}
\caption{{Optimal navigation in a Taylor-Green flow.} 
(a) The Taylor-Green flow map, where the colour codes for the intensity and the black arrows indicate the direction.
The green curve indicates Zermelo's optimal trajectory at vanishing noise connecting the initial point $\bm{r}_0 = -(\ell/2) \hat{\bm e}_x$ and the target $\bm{r}_T=\bm 0$, marked with a green and magenta circle, respectively. The red and cyan lines show two representative stochastic trajectories following respectively the Optimal and Straight policies at $\gamma = 0.7$ and $\Pe = 400$.
(b) The optimal control obtained by solving Eqs.~(\ref{eq:HJB}) and (\ref{eq:optcontrol}) for same parameters as in panel (a).
(c) Schematic defining the quantities used for the implementation of the semi-autonomous navigation strategies.
The red dot indicates the active particle position $\bm{r}$ and the red arrow its heading direction $\bm{\hat{u}}$. 
$\bm{r}_c$ (red cross) marks the closest point to $\bm r$ on the optimal trajectory, while the two vectors $\bm{\hat{n}}$ and $\bm{\hat{t}}$ are respectively normal and tangent to the optimal trajectory in $\bm{r}_c$.} 
\label{fig:1}
\end{figure*}

\section{Results}
\subsection{Supervised optimal navigation} 
We consider an overdamped self-propelled particle moving at a fixed speed $v_0$ in presence of a stationary force field $\bm{f}(\bm{r})$, which may in general include a contribution due to advection by the solvent flow velocity, and translational diffusion with diffusivity $D$.  For simplicity, we set the friction coefficient (and mobility) to unity. The position $\bm r$ of the self-propelled particle obeys the following stochastic differential equation
\begin{equation}
    \dot{\bm r}  = v_0 \bm{\hat{u}} +\bm{f}(\bm{r}) + \sqrt{2D}\,\boldsymbol{\xi} \ , 
    \label{eq:motion}
\end{equation}
where $\bm{\hat{u}}$ is the unit vector setting the direction of self-propulsion, 
and $\boldsymbol{\xi}$ is a Gaussian white noise vector whose components have zero mean and unit variance.
Within this setting, the only degree of freedom accessible to the self-propelled particle for navigation is its orientation $\bm{\hat{u}}$.
For the sake of presentation, we will first assume a full control over $\bm{\hat{u}}$, either from external sources or by the particle itself,
while this constraint will be relaxed later.

We now want to determine the optimal navigation protocol for the particle to reach a {\em target} position $\bm r_T$ given a force profile and an initial position $\bm r_0$. Following standard techniques that invoke the backward Fokker-Planck equation~\cite{Risken1996}, the {\em mean first-passage time} (MFPT) ${\cal T}(\bm r)$ 
to reach $\bm r_T$ while starting initially at $\bm r$ can be found as the solution of the following equation 
(see {Methods} for a derivation):
\begin{equation}
        -v_0 |\bm{\nabla}{\cal T}| + \bm{f}(\bm{r}) \cdot \bm{\nabla}{\cal T} + D\bm{\nabla}^2 {\cal T} = -1 \ ,
    \label{eq:HJB}
\end{equation}
which is to be solved subject to the boundary condition ${\cal T}(\bm r_T) = 0$. Here, the prescription for optimal control comprises tuning the orientation according to the the following rule
\begin{equation}
    \bm{\hat{u}}_{\rm opt}(\bm r) = -\bm{\nabla}{\cal T}/|\bm{\nabla}{\cal T}| \ ,
    \label{eq:optcontrol}
\end{equation}
at every point in space. 
We note that Eqs.~(\ref{eq:HJB}) and (\ref{eq:optcontrol}) provide strategies which, by design, lead to the fastest possible trajectories on average as can be seen in the vanishing diffusivity limit where Zermelo's solution~\cite{zermelo} is recovered. Hereafter, we will refer to the strategy that corresponds to the solution of Eqs.~(\ref{eq:HJB}) and (\ref{eq:optcontrol}) as the \emph{optimal policy} (OP), and the $D = 0$ optimal path as the \emph{Zermelo path}.

We now illustrate OP by considering a simple but nontrivial setup in which a self-propelled particle navigates in the two dimensional plane spanned by the unit vectors $\{\hat{\bm e}_x,\hat{\bm e}_y\}$ between neighbouring stationary points of a Taylor-Green vortex flow 
(see the colour map and black arrows in Fig.~\ref{fig:1}a).
This configuration corresponds to $\bm{f}(\bm{r}) = v_f [\cos(ky) \sin(kx)\,\hat{\bm e}_x -\cos(kx) \sin(ky)\,\hat{\bm e}_y]$ with $k =2 \pi / \ell$ and $\ell$ being the characteristic length scale of the flow. Rescaling space and time as $\bm r \to \ell \bm r$ and $t \to \ell t / v_0$, 
the dynamics~\eqref{eq:motion} is characterized by only two nondimensional parameters: 
the ratio between flow intensity and self propulsion $\gamma \equiv v_f / v_0$, and the P\'eclet number $\Pe= \ell v_0 / D$. Here, we will focus only on cases where the self-propulsion is always stronger than the flow, namely, $0 \leq \gamma \leq 1$.

\begin{figure*}[t!]
\includegraphics[width=\linewidth]{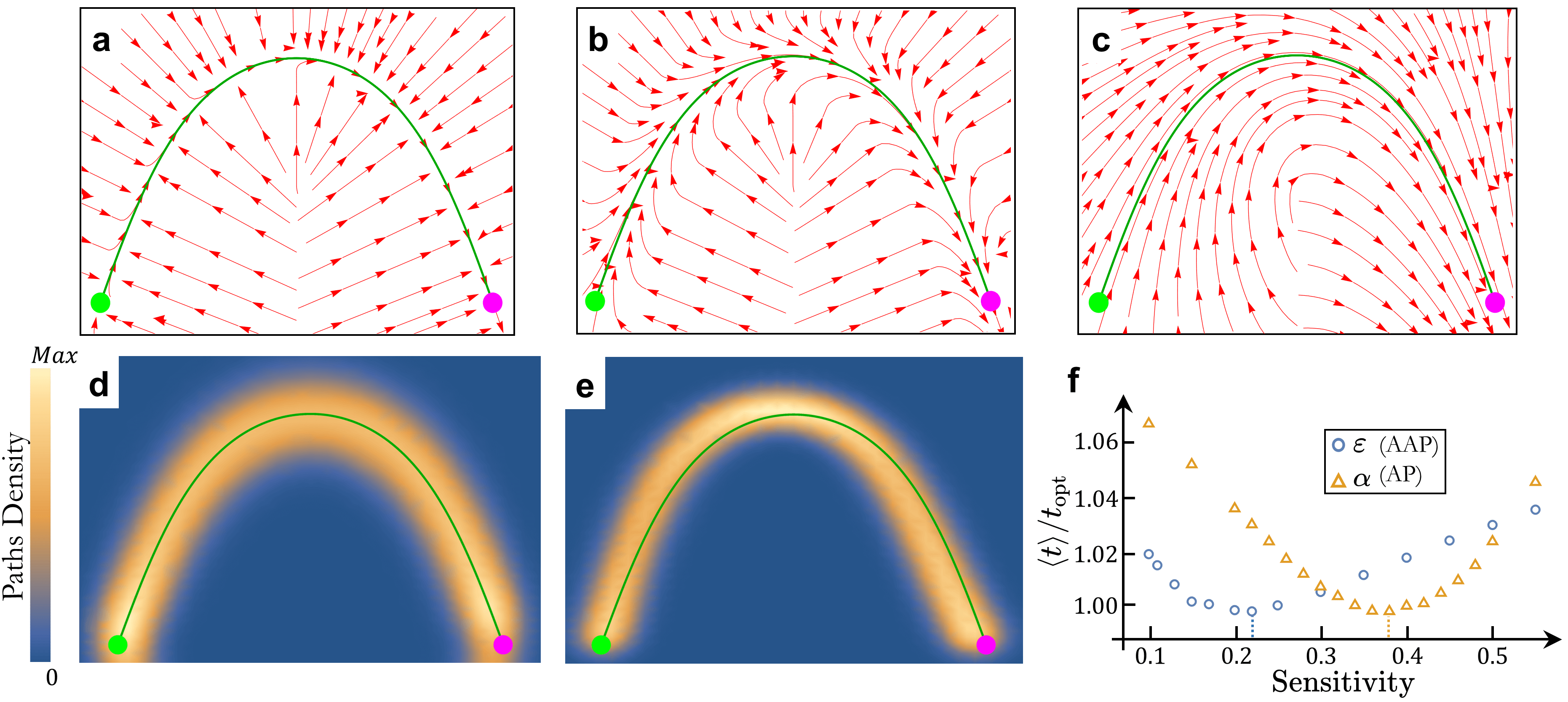}
\caption{{The semi-autonomous navigation policies.} 
(a)-(c) The control $\bm{\hat{u}}$ obtained for the AAP from Eqs.~(\ref{eq:policy_general}) and (\ref{eq:stick}) in the Taylor-Green flow at $\gamma=0.7$ with $\varepsilon = 0.05$(a), 0.2(b) and 1(c).
(d,e) Heat maps of $10^3$ stochastic trajectories obtained from numerical simulations of the OP(d) and AAP(e). 
The solid green lines in (a)-(e) represent the $D = 0$ Zermelo path. 
(f) Example of curves leading to the determination of the optimal sensitivities for the AP and AAP ($\gamma = 0.7$ and $\Pe = 400$) with $\avgt$ and $t_{\rm opt}$ respectively denoting the mean arrival time and the optimal travel time in absence of noise. The optimal values of $\alpha$(AP) and $\varepsilon$(APP) are indicated by the vertical dashed lines.
Note the small range of values for the normalized mean arrival time, showing the robustness of the two policies. Here, both curves are normalized by their minimum value.} \label{fig:2}
\end{figure*}

With the setup shown in Fig.~\ref{fig:1}a, the most direct route between the departure and arrival points requires travelling counter-flow all the way. Consequently, the straight path becomes increasingly disadvantageous as the flow amplitude grows, such that for $\gamma\gtrsim 0.4$ the Zermelo path 
makes use of the flow profile and takes a bell-shaped curve (see the green line in Fig.~\ref{fig:1}a). 
For finite P\'eclet number ($D \neq 0$), such a feature is moreover consistent with the control map provided by OP, since the latter generally orients the self-propulsion away from the straight path (Fig. \ref{fig:1}b). Simulations of the Brownian dynamics~\eqref{eq:motion} with the control map solving Eqs.~(\ref{eq:HJB}) and (\ref{eq:optcontrol}) indeed reveal that the OP trajectories tend to remain close to the Zermelo path for a broad range of P\'eclet number values, provided that the flow strength and particle self-propulsion dominate over fluctuations (see e.g. the density map in Fig. \ref{fig:2}d and Methods for more details on the stochastic dynamics simulations).\\

\subsection{Semi-autonomous optimal navigation}
The above observations suggest that optimized navigation in the finite P\'eclet number regime may be achievable using only the local information of the relative positions of the stochastic swimmer and the Zermelo path, as opposed to the OP which requires a global control map $\sim\bm{\nabla} {\cal T}(\bm r)$.

We now explore a number of such local strategies and probe their efficiencies in comparison with OP. For a given particle position $\bm r$, we define $\bm r_c \equiv \min_{\bm r'} |\bm r - \bm r'|$ as the corresponding closest point on the Zermelo path.
Moreover, we suppose that the latter is smooth and can be parametrized by the moving frame $\{\hat{\bm t},\hat{\bm n}\}$,
of tangent and normal vectors, with $\hat{\bm t}$ heading towards the target as shown in Fig. \ref{fig:1}c. Assuming that the swimmer is able to measure its relative position to the Zermelo path, it can regulate it by steering its self propulsion direction $\bm{\hat{u}}$ via the following rule
\begin{equation}
    \bm{\hat{u}} \cdot \bm{\hat{n}} = {\cal G}(\Delta r,\bm f(\bm r))  \ ,
    \label{eq:policy_general}
\end{equation}
where $\Delta r \equiv (\bm{r}-\bm{r_c}) \cdot \bm{\hat{n}}$ and the function ${\cal G} \in [-1,1]$ depends on the amount of information available to the swimmer.
As the rhs of Eq.~\eqref{eq:policy_general} depends on $\bm r$ solely through $\Delta r$ and the external flow $\bm f$, 
it defines a class of optimal navigation policies relying only on the swimmer's local knowledge of its environment.

In the simplest case where the swimmer can only determine the direction $\hat{\bm n}$ to the Zermelo path (from its current position),
it can choose to keep a constant angle $\alpha$ between its self-propulsion direction and $\hat{\bm n}$.
Such \emph{aligning policy} (AP) corresponds to a protocol ${\cal G}_{\rm AP} = \pm\cos(\alpha)$, where the $\pm$ sign ensures that $\bm{\hat{u}} \cdot \hat{\bm t} \ge 0$.

Although AP allows the swimmer to remain in the vicinity of the Zermelo path, it also slows it down as it imposes a finite angle between $\bm{\hat{u}}$ and $\hat{\bm t}$ even for (arbitrarily) small separations. 
For swimmers able to evaluate their relative distances to the Zermelo path, AP can thus be refined by allowing ${\cal G}$ to depend on $\Delta r$. This defines the \emph{adaptive aligning policy} (AAP). Here, we choose, for simplicity, a piecewise linear form for ${\cal G}(\Delta r)$, namely
\begin{equation}
    {\cal G}_{\rm AAP}(\Delta r) = 
    \begin{cases}
        +1 & \Delta r < -\varepsilon \\
        -\Delta r/\varepsilon & |\Delta r| < \varepsilon \\
        -1 & \Delta r > \varepsilon 
    \end{cases} ,
    \label{eq:stick}
\end{equation}
where the parameter $\varepsilon$ sets a cut-off scale above which the stochastic particle points normally to the Zermelo path.

\begin{figure*}[t!]
\includegraphics[width=\linewidth]{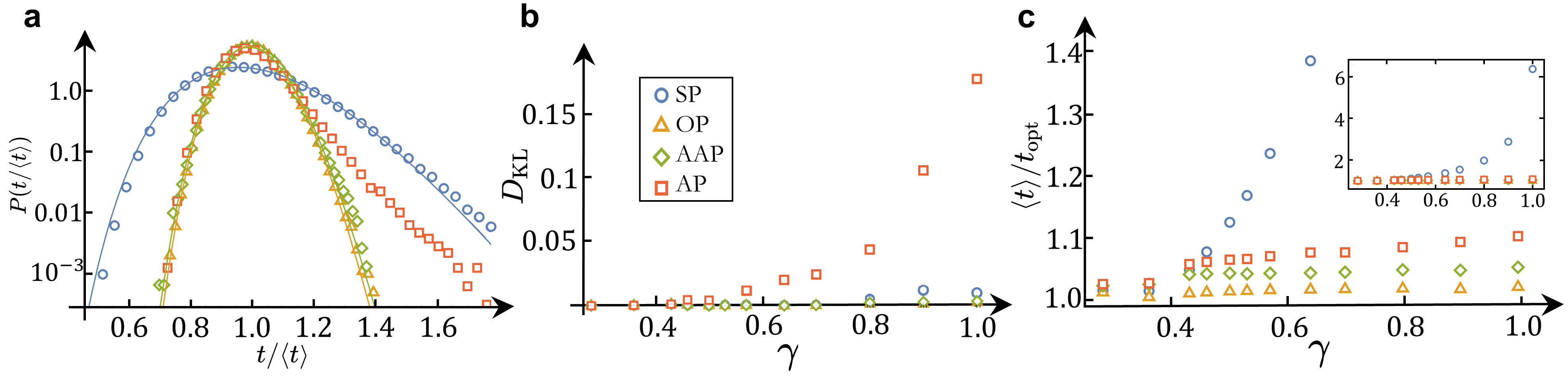}
\caption{{Comparison of the performances of the different policies.} 
(a) Arrival time distribution as a function of the normalized time $t / \avgt$ for the four navigation strategies:
straight policy (blue circles), optimal policy (yellow triangles), adaptive aligning policy (green diamonds) and aligning policy (red squares). 
(b) Kullback-Leibler divergence $D_{\rm KL}$ between the numerical and theoretical (Eq.~\eqref{eq:Inv_G}) distributions as a function of the relative flow strength $\gamma$.
(c) Mean arrival time (normalized by the noiseless optimal time $t_{\rm opt}$) as a function of $\gamma$.
Inset: Zoom-out showing the larger values taken by SP. In (a-c) the symbols show the data obtained from Langevin simulations, while in (a) the solid lines indicate the theoretical prediction~\eqref{eq:Inv_G} calculated with $\avgt$ and $\sigma$ obtained from the numerical data. 
All data in (a)-(c) are averaged over $10^5$ independent trajectories.} \label{fig:3}
\end{figure*}

The parameters $\alpha$ and $\varepsilon$ introduced above essentially play the role of sensitivities for AP and AAP, respectively.
Their optimal values (that minimize the mean travel time in this case) cannot be selected {\it a priori}, and need to be determined empirically.
However, the existence of such optimal values can be intuitively understood from the control maps obtained from Eq.~\eqref{eq:stick} for various $\varepsilon$ values (with the generalization to AP being straightforward). As shown in Figs.~\ref{fig:2}a-c, exceedingly small $\varepsilon$ values force the swimmer to mostly point normally to the Zermelo path, while for excessively large $\varepsilon$ stochastic trajectories are less efficiently confined and can visit less favourable flow regions. Therefore, it is natural to expect an intermediate value of $\varepsilon$ providing the optimal trade-off between efficient confinement and tangential motion along the Zermelo path.

This heuristic picture is confirmed by numerical simulations showing that the mean arrival time $\langle t \rangle$ indeed exhibits a minimum at a value $\varepsilon = \varepsilon_{\rm opt}$ (see Fig. \ref{fig:2}f). We moreover note that $\langle t \rangle$ varies relatively little with $\varepsilon$, such that in practice the policy implementation does not require a fine tuning of this parameter. The heat map of trajectories obtained from simulations of AAP at optimal $\varepsilon$ shows that they globally follow the Zermelo path (Fig.~\ref{fig:2}e), similarly to the OP case. Contrary to OP, however, they are not distributed symmetrically with respect to the desired path due to a non-zero transverse component of the flow (see Fig.~\ref{fig:1}a). Better conformity to the OP results thus requires additional features such as a function $\cal G$ in 
Eq.~\eqref{eq:policy_general} that depends explicitly on the local flow field $\bm f(\bm r)$. Here, we restrict to the case where the swimmer is unaware of the local flow structure around it and will address such more elaborate policies in a forthcoming publication~\cite{nextpaper}.\\

\subsection{Performance assessment of the navigation policies}
We now compare the performances of the two policies introduced above (AP and AAP) with that of OP by simulating Eq.~\eqref{eq:motion} 
with the controls defined by~(\ref{eq:optcontrol}) and (\ref{eq:policy_general}) in the Taylor-Green flow setup (Fig.~\ref{fig:1}a).
To illustrate the relevance of nontrivial policies, we moreover consider the \emph{straight policy} (SP) for which the swimmer always points towards the direction of the target regardless of its current position (see the cyan curve in Fig.~\ref{fig:1}a for a representative trajectory).

\begin{figure*}[t!]
\includegraphics[width=\linewidth]{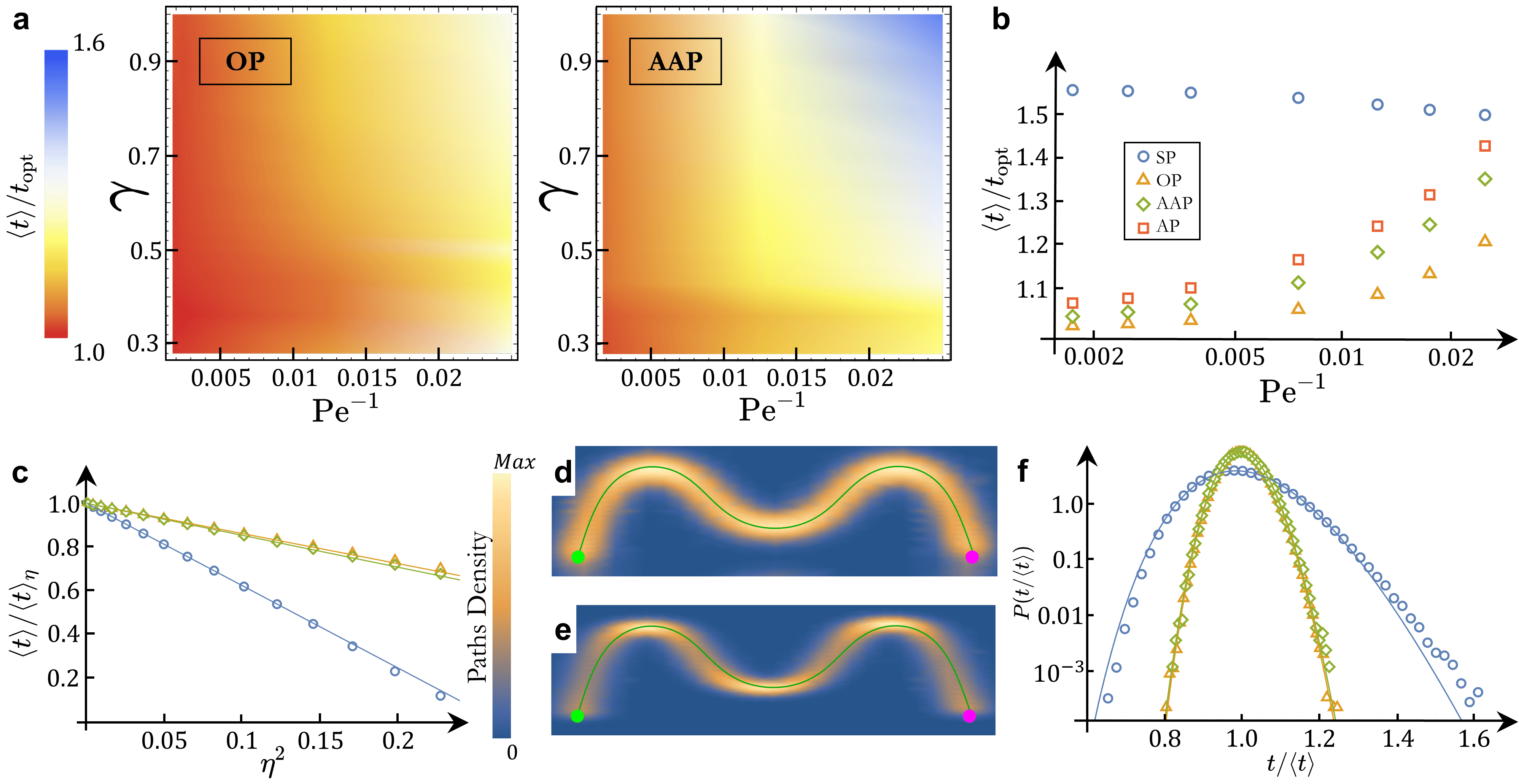}
\caption{{Robustness of the navigation policies.} 
(a) Colour maps showing the performances of OP (left) and AAP (right) as functions of the relative flow amplitude $\gamma$ and the P\'eclet number $\Pe$. (b) Normalized mean arrival time as a function of the P\'eclet number at $\gamma = 0.7$ for the four policies (horizontal axis in log scale). (c) Inverse of the mean arrival time as a function of the rotational noise strength $\eta^2$ at $\Pe=400$ and $\gamma = 0.7$ for OP, AAP and SP (caption is the same as in (b)). Here the data is normalized by the value $\avgt$ measured at $\eta = 0$. Solid lines show linear fits in the small $\eta$ regime. (d), (e) Heat maps of $10^3$ stochastic trajectories obtained from numerical simulations at $\gamma=0.7$ and $\Pe=400$ of OP and AAP, respectively. In both panels  the green curve shows the Zermelo path connecting the initial point $\bm{r}_0=-(3\ell/2) \hat{\bm e}_x$ (green circle) and the target $\bm{r}_T=\bm 0$ (magenta circle). (f) Arrival time probability distributions as functions of $t / \avgt$ corresponding to (d,e) as well as for the SP case under similar conditions; the caption is the same as (b). Solid lines show the theoretical curves obtained similarly to Fig.~\ref{fig:3}(a). All data in (a)-(c) and (f) are averaged over $10^5$ independent trajectories.
} \label{fig:4}
\end{figure*}

We first work at fixed $\Pe=400$ and vary the relative flow amplitude $\gamma \in [0,1]$. Figure~\ref{fig:3}a shows the arrival time distributions $P(\tau)$ with $\tau \equiv t / \langle t \rangle$ for each of the four policies (OP, AP, AAP, and SP) at $\gamma = 0.7$.
Remarkable overlap between the OP and AAP distributions can be observed. We moreover find that they are both well described by a so-called inverse Gaussian distribution of the form
\begin{equation}
    P(\tau) = \sqrt{\frac{\avgt^2}{2\pi\sigma^2 \tau^3}} \exp\biggr[-\frac{\avgt^2(\tau-1)^2}{2\sigma^2\tau}\biggr] \ ,
    \label{eq:Inv_G}
\end{equation}
with variance $\sigma^2$. To quantify this correspondence we furthermore calculate the Kullback-Leibler divergence 
$D_{\rm KL} = \langle \ln [P_{\rm num}(\tau) / P(\tau)] \rangle_{P_{\rm num}}$ between the numerically obtained distribution $P_{\rm num}$ and the prediction of Eq.~\eqref{eq:Inv_G}, with $\avgt$ and $\sigma$ determined from the data. As shown in Fig.~\ref{fig:3}b, $D_{\rm KL}$ remains almost zero for both OP and AAP over a wide range of $\gamma$ values, highlighting the robustness of~\eqref{eq:Inv_G}. As the inverse Gaussian corresponds to the First Passage Time distribution of a driven Brownian particle in one dimension \cite{folks}, Eq.~\eqref{eq:Inv_G} is closely related to the confinement of the OP and AAP trajectories along the Zermelo path as shown in Figs.~\ref{fig:2}d-e. In fact, for both OP and AAP the loss of correspondence with the inverse Gaussian coincides with the regime of strong fluctuations that prevent the swimmers from being efficiently guided along the Zermelo path (details in SI). 

In contrast, for sufficiently large flow amplitudes the simpler aligning policy shows arrival time distributions that do not follow the inverse Gaussian law (red squares in Figs.~\ref{fig:3}a,b). These distributions indeed exhibit a crossover from inverse Gaussian-like behaviour at $\tau < 1$ to an exponential decay at $\tau > 1$ with a characteristic time $\tau_{\rm AP}$ that is systematically larger than the value $2\sigma^2/\avgt^2$ predicted by Eq.~\eqref{eq:Inv_G}. As detailed in the SI, these deviations from the inverse Gaussian law correspond to asymmetric trajectory distributions around the Zermelo path, similarly to the case observed for OP and AAP in the large noise regime. In particular, for the AP case most trajectories land on the left of the target such that in most cases the swimmer has to navigate counter-flow in the last part of its journey.

Finally, SP shows arrival time distributions globally compatible with the inverse Gaussian~\eqref{eq:Inv_G}, with only slight deviations at large flow amplitudes (see blue circles in Figs.~\ref{fig:3}a,b), indicating that in this case too trajectories are nearly one dimensional. As SP trajectories are mostly oriented against the flow, they are characterized by a lower effective drift on average, resulting in a larger ratio $\sigma / \avgt$.

\begin{figure*}[t!]
\includegraphics[width=\linewidth]{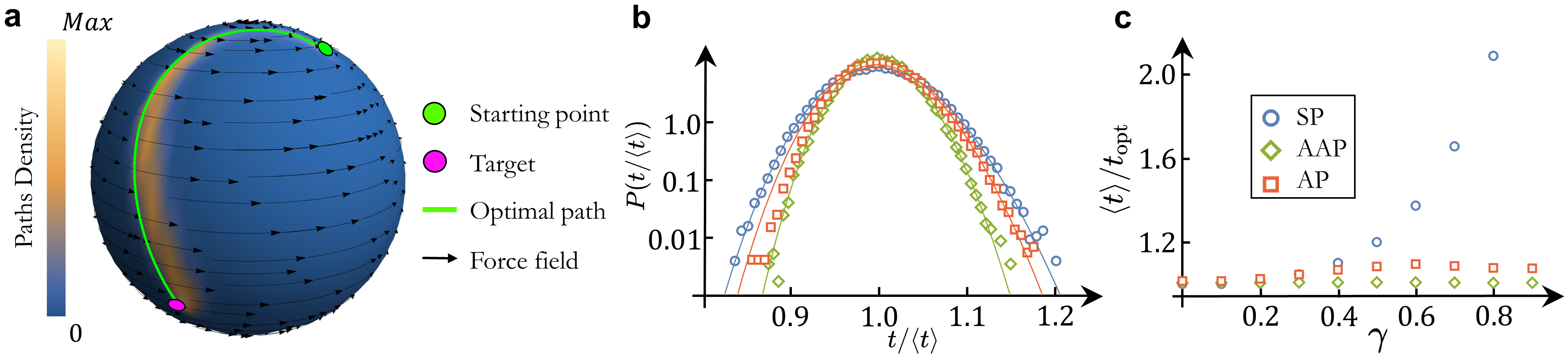}
\caption{{Optimal navigation on a sphere.} 
(a) Illustration of the chosen setup. The black arrows indicate the direction of the flow while the solid green curve connecting the points $\bm{r}_T = (2\pi/3) \hat{\bm e}_\theta + (3\pi/2) \hat{\bm e}_\phi$ (magenta circle) and 
$\bm{r}_0 = (\pi/6) \hat{\bm e}_\theta$ (green circle) corresponds to the Randers geodesic at $\gamma = 0.5$.
Here the colour map shows the trajectory distribution for AAP. 
(b) The arrival time distribution as a function of $t /\avgt$ for straight (blue circles), adaptive aligning (green diamonds) and aligning (red squares) policies at $\gamma = 0.5$.
(c) Mean arrival time as a function of $\gamma$ for the three policies; the caption is the same as (b). 
In (a-c) the value of the P\'eclet number is set to $\Pe=10^3$. All data in (b,c) are averaged over $10^5$ independent trajectories.
} \label{fig:5}
\end{figure*}

As customary in this context, we compare the performances of the navigation policies by measuring the mean arrival time to reach the target $\langle t\rangle$~\footnote{Other measures of performances which take into account additional features of the arrival time distribution can be defined. 
As detailed in the SI, we find that while those allow for more refined evaluation of the efficiency, they lead to similar results to those obtained from the mean arrival time $\avgt$.}, here normalized using the optimal value $t_{\rm opt}$ at $D = 0$ obtained from Zermelo's solution.
The corresponding results shown in Fig. \ref{fig:3}c reveal that, naturally, OP performs the best with the mean arrival times always remaining higher than $t_{\rm opt}$ by only a few percent. Conversely, the trivial SP performances strongly deteriorate as $\gamma$ increases.
For small flow amplitudes where the Zermelo path is almost straight ($\gamma \lesssim 0.4$) SP performs similarly to OP with $\langle t \rangle \gtrsim t_{\rm opt}$, whereas for sufficiently large $\gamma$ values it exhibits mean arrival times reaching five to six times $t_{\rm opt}$ (see the inset of Fig.~\ref{fig:3}c). Despite the presence of fluctuations, the performances of the policies are thus primarily set by their ability to make efficient use of the stationary flow profile. This feature is moreover illustrated by both AP and AAP, which show mean arrival times no more than $10\%$ higher than that of OP, regardless of the relative flow amplitude. As expected, in the non-trivial cases ($\gamma > 0.4$) the performances of the different strategies reflect the amount of information they require for navigation, such that, in the order of increasing efficiency, one finds SP, AP, AAP and OP.\\

\subsection{Robustness of the new protocols}
The above analysis shows that AAP displays arrival time statistics similar to that of OP. Both AP and AAP moreover exhibit performances comparable to OP, despite them relying only on local information. We now assess the generality of these results, focusing on AAP, by discussing more general situations with different model parameters and evaluation setups.

\noindent {\it The role of $\Pe$.}
The two colour maps of Fig.~\ref{fig:4}a show how the ratio $\langle t \rangle / t_{\rm opt}$ varies with the P\'eclet number and relative flow amplitude $\gamma$ for OP and AAP. In agreement with previous results, $\langle t \rangle / t_{\rm opt}$  for OP does not significantly depend on $\gamma$ while we observe a slight increase with decreasing $\Pe$. The AAP case, on the other hand, exhibits two distinct regimes.
At small flow strengths ($\gamma \lesssim 0.4$), $\langle t \rangle / t_{\rm opt}$ remains nearly constant upon varying $\Pe$ such that the AAP performances are not significantly altered by the amplitude of noise.
Conversely, at larger $\gamma$ values where the Zermelo path is more curved, the mean arrival time is more affected by translational noise.
As shown in Fig~\ref{fig:4}b for $\gamma = 0.7$, all non-trivial strategies show a slight decrease in performance as $\Pe$ is lowered, whereas SP becomes slightly more favourable upon increasing the noise, since in this case stronger fluctuations lead the swimmer to visit less unfavourable flow regions.

\noindent {\it Misalignment of self-propulsion.}
We have so far assumed full control over the self-propulsion orientation $\bm{\hat{u}}$. In reality, however, $\bm{\hat{u}}$ is subject to fluctuations---e.g. due to rotational Brownian motion or inaccuracies in the evaluation of the desired direction---which affect the performances of the policies using it as a control. To model the effect of rotational noise, we applied random rotations $\bm{\hat{u}} \to {\cal R}(\beta)\bm{\hat{u}}$ to the controls~(Eqs. (\ref{eq:optcontrol}) and (\ref{eq:policy_general})), where the angle $\beta$ was sampled from a uniform distribution in $(-\eta\pi, \eta\pi]$; we present the corresponding results in Fig.~\ref{fig:4}c.
For small $\eta$ values, we find that the inverse of the mean arrival time normalized by its value at $\eta=0$, $\avgt/\avgt_\eta$,
decays linearly with $\eta^2$ with a policy-dependent slope. In particular, OP and AAP show similar trends and appear to be much more robust to the effect of rotational noise than the SP case, as for the latter the mean arrival time has increased by a factor $10$ at $\eta = \tfrac{1}{2}$, while the corresponding drop in performance for OP and AAP is about $40\%$. We moreover show in Methods how the scaling $\avgt/\avgt_\eta - 1 \sim \eta^2$ can be derived from a simple argument in the limit of small noises, and how it is related to the quasi-one dimensional nature of the problem. This scaling is thus not expected to hold for large noises, as suggested by the deviations from the linear decay observed for the largest $\eta$ values in Fig.~\ref{fig:4}c.

\noindent {\it Complex navigation tasks.}
Increasing the distance between the initial and target points in the Taylor-Green flow allows us to design more complex paths. Upon translating the initial swimmer position along $\hat{\bm e}_x$, as shown in Figs.~\ref{fig:4}d and~\ref{fig:4}e, both OP and AAP lead to trajectories focused around the Zermelo path on average. Consequently, the corresponding arrival time distributions remain well characterized by the inverse Gaussian law (Figs.~\ref{fig:4}f). The performances of AAP and OP are moreover stable upon increasing the total travel distance, with mean arrival times $\avgt$ not higher than $t_{\rm opt}$ by more than a few percent. \\

\subsection{Optimal navigation on a manifold}
We next show how AP and AAP navigation protocols are applicable to motion on curved landscapes. Self-propelled motion on curved surfaces has recently earned growing attention both at individual~\cite{fily2016active,apaza2018SoftMat,villareal2018,piro} and collective~\cite{sknepnek2015PRE,sknepnek2018,lin2020,napoli2020PRE,shankar2021topological} levels.
As stochastic motion taking place on a generic Riemannian manifold involves multiplicative noise, solving the corresponding MFPT equation~\eqref{eq:HJB} requires advanced computational techniques~\cite{dziuk2013}, which will introduce additional challenges for determining the stochastic optimal control~\eqref{eq:optcontrol}. On the other hand, Zermelo's approach was recently generalized to self-propelled motion on curved surfaces using a mapping to Finsler geometry~\cite{piro}. The corresponding noiseless optimal path, also known as the \emph{Randers geodesic}, can then straightforwardly be used to extend AP and AAP policies to non-Euclidean spaces.

For the sake of illustration, let us consider the case of active motion on a sphere in the presence of a unidirectional flow ${\bm f}(\theta,\phi) = v_f \sin \theta \,\hat{\bm e}_\phi$,
where $\theta$ and $\phi$ respectively denote the polar and azimuthal angles in the spherical coordinate system.
As shown in Fig.~\ref{fig:5}a (see the black arrows), this flow---which is characterized by a pair of vortices at the poles and is maximum at the equator---generally leads to non-trivial Randers geodesics (solid green line) between two arbitrary points on the sphere.

Simulating the counterpart of the Langevin equation~\eqref{eq:motion} on the sphere (details in Methods) at fixed $\Pe = 10^3$, 
we are able to compare the performances of AP, AAP and SP. The corresponding arrival time distributions are shown in Fig.~\ref{fig:5}b.
As for the Taylor-Green flow in flat space, we find that they are all in good agreement with the inverse Gaussian law~\eqref{eq:Inv_G}.
In the AP case, this result is probably due to the rather large value of $\Pe$ chosen for convenience, which allows all policies to exhibit trajectories well distributed around a one dimensional path. Figure~\ref{fig:5}c moreover shows that at small flow strengths all policies perform similarly with $\langle t \rangle \gtrsim t_{\rm opt}$, while for larger $\gamma$ values leading to more complex Randers geodesics SP becomes increasingly disadvantageous. On the contrary, both AP and AAP always remain close to optimality, as they exploit the information of the noiseless optimal path. 
 \\

\section{Discussion}
We have introduced a class of policies that allow for semi-autonomous optimal navigation of microswimmers in complex and noisy environments. 
These policies rely on the swimmer knowledge of its local environment, such as its relative position and distance to a desired path, and their performances were found to improve with the amount of information available to the swimmer. In particular, extensive numerical simulations reveal that the adaptive aligning policy (for which the self-propulsion orientation varies with the distance to the Zermelo path) shows performances comparable to that obtained from the optimal MFPT control. The two strategies lead to statistically similar trajectories and nearly identical inverse Gaussian arrival time distributions. Our analysis moreover shows that the best performing strategies are also the most robust to environmental changes, such as stronger translational diffusion and the introduction of rotational noise. Finally, it was shown that the newly introduced navigation strategies have the additional advantage of being easily applicable to the problem of optimal navigation on curved surfaces. In an illustrative example on spherical geometry, the semi-autonomous navigation strategies were once again found to perform significantly better than the trivial one consisting of pointing straight to the target.

Although the analysis carried out here focused on the problem of travel time optimization, the policies we proposed are fully determined by the noiseless optimal path. Therefore, they are straightforwardly generalizable to a broader class of optimization problems like energy dissipation or fuel consumption minimization. 
Furthermore, even though the new policies show remarkable performances as compared to OP, some differences persist. In particular, while trajectories following OP are symmetrically distributed with respect to the Zermelo path, this is not the case for AP and AAP (Figs.~\ref{fig:2}(d,e)). Our ongoing work investigates how further improvement could be reached by designing policies based on the ability of some swimmers to adapt their swimming direction according to the local flow field~\cite{saha2014PRE,Palacci2006sciadv}.

Lastly, we note that the constraint of imposing a constant sensitivity (represented by the parameters $\alpha$ and $\varepsilon$) 
throughout the swimmer motion might restrict the performances of the policies. While making these parameters explicitly space dependent would break the semi-autonomous nature of the policies, determining the functions $\alpha(\bm r)$ or $\varepsilon(\bm r)$ is certainly a simpler problem than calculating the full optimal control map of the swimmer orientation. Therefore, the framework presented here could serve as a basis for reinforcement learning based approaches applied to complex navigation problems~\cite{colabrese,Schneider2019EPL,muinos2021ScienceR}.\\


\section{Methods}
\textbf{The optimal control for an active particle in a stationary flow.}
The scope of this section is to provide a derivation of the mean first passage time equation and the corresponding optimal control (Eqs. \eqref{eq:HJB} and \eqref{eq:optcontrol} in the main text). The derivation relies on standard techniques presented in Ref.~\cite{Risken1996}.

First of all, we aim to derive a policy that minimizes the mean arrival time at the target given an initial position $\bm x$. 
Let us therefore denote ${\cal T}(\bm x)$ 
as the mean travel time. 
Now, we consider the joint probability $p(\bm{y},t|\bm{x},0)$ that at time $t$ the swimmer is at position $\bm{y}$, given that it started from $\bm{x}$ at $t=0$. This probability obeys the backward Fokker-Planck equation (from \eqref{eq:motion}):
\begin{equation}
    -\partial_t p + (v_0 {\bm{\hat{u}}}(\theta) +  {\bm f}({\bm x}))\cdot \bm{\nabla}p+ D\nabla^2p = 0 \ ,
    \label{eq:FP_backward}
\end{equation}
where the gradients are taken with respect to $\bm{x}$.
Moreover, $p$ is related to ${\cal T}$ via
\begin{equation*}
    {\cal T}(\bm{x})=\int  {\rm d}t {\rm d} \bm{y}\, p(\bm{y},t|\bm{x},0)  ,
\end{equation*}
such that we obtain
\begin{equation}
    (v_0 {\bm{\hat{u}}}(\theta) + {\bm f}({\bm x}))\cdot \bm{\nabla}{\cal T} + D\nabla^2{\cal T} = -1 \ .
    \label{eq:Feq}
\end{equation}
The optimal choice for the heading direction $\theta$ (our control parameter) can be obtained by taking the variational derivative of both sides of \eqref{eq:Feq} with respect to $\theta$ itself, leading to
\begin{equation}
    \tan\theta=\frac{\partial_y {\cal T}}{\partial_x {\cal T}} \iff \bm{\hat{u}}(\theta)= -\frac{\bm{\nabla}{\cal T}}{|\bm{\nabla}{\cal T}|} \ ,
    \label{eq:optcontrol2}
\end{equation}
which corresponds to Eq. \eqref{eq:optcontrol} in the main text.  Putting together Eqs. \eqref{eq:Feq} and \eqref{eq:optcontrol2}, 
we get the MFPT equation of our problem (Eq.\eqref{eq:HJB} in the main text).\\

\noindent \textbf{Numerical methods.}
The Brownian dynamics simulations of Eq. \eqref{eq:motion} have been performed using an Euler-Mayurama scheme with a time step $dt=10^{-3}$. 
We have verified that the selected time step is sufficiently small, such that the results presented here do not depend on its value. 
In all our simulations a given run ends when the active particle is within a distance $\delta r=0.025 \ell$ from the target. 
We have also checked that the choice of the disk radius $\delta r$ does not significantly influence the results as long as the thermal fluctuations length scale is kept relatively small, i.e. 
$\sqrt{2D dt}\ll \delta r \ll \ell$.\\

\noindent \textbf{Details on the implementation of the protocols.}
Here we provide details regarding the implementation of the navigation strategies presented in the main text.

The MFPT equation~\eqref{eq:HJB} was numerically solved using the Finite Elements Method implemented in the NDSolve routine of Wolfram Mathematica 12.3.1~\cite{mathematica}. The Optimal policy was then implemented from the corresponding solution~\eqref{eq:optcontrol} by discretizing the simulation domain on a square grid of step $l=0.01$, and assigning to each box the optimal control orientation $\bm{\hat{u}}_{\rm opt}(\bm r_b)$, with $\bm r_b$ being the position of the centre of the box. In the stochastic simulations, the swimmer following OP was then aligning its direction of motion with the orientation associated with its current position on the grid.

Both AP and AAP rely on the evaluation of the point ${\bm r}_c$ on the Zermelo path closest to the particle position $\bm r$. For numerical efficiency, the Zermelo path was thus discretized and the distance between the particle and the curve was calculated from the positions of the mid-point of each segments. In all simulations the initial particle orientation was taken to be equal to the one prescribed by the Zermelo solution.\\

\noindent \textbf{The scaling of the mean arrival time with rotational noise.}
Here, we show how the linear scaling of the mean arrival time $\avgt_\eta$ with the rotational noise strength $\eta^2$ shown in Fig.~\ref{fig:4}c can be understood from an effective one dimensional model of driven Brownian motion.

Assuming that the particle remains in the vicinity of the mean path and neglecting effects due to the curvature of the latter, we consider the following dynamics
\begin{equation}
    \dot{r}_\| = v_\|(r_\|) \cos\theta + \sqrt{2D}\xi_\| ,
    \label{eq:motion_x}
\end{equation}
where the subscript $\|$ stands for quantities projected along the mean path. The first term on the rhs of~\eqref{eq:motion_x} thus accounts for the total mean velocity of the particle along the mean path, which includes the combined effects of flow and self-propulsion. In general, the angle $\theta$ obeys a nontrivial and policy dependent dynamics. However, in the limit of small $\eta$ and $D$
where the particle remains close to the mean path, we approximate $\theta$ as a Gaussian noise with zero mean and variance $\propto\eta^2$.
Expanding the cosine and performing an average over the noises we thus obtain
\begin{equation*}
    \langle \dot{r}_\| \rangle \simeq v_\|(r_\|)(1 - \kappa\eta^2) ,
\end{equation*}
with $\kappa>0$ being a constant that depends on the navigation details, e.g. the protocol used. In the one-dimensional approximation and assuming that $v_\|$ varies little with $r_\|$, the mean travelling time to reach an absorbing barrier at distance $L$ scales as $\avgt_\eta \sim L/\langle\dot{r}_\|\rangle$, such that we obtain
\begin{equation*}
    \avgt/\avgt_\eta \simeq 1 - \kappa\eta^2 \ ,
\end{equation*}
which corresponds to the scaling observed in Fig.~\ref{fig:4}c.\\

\noindent \textbf{Langevin simulations on the sphere.}
To describe the motion of an overdamped particle on the sphere, the Langevin equation~\eqref{eq:motion} has to be adjusted to take into account the multiplicative noise induced by the space curvature. Namely, it is given by
\begin{equation}
    \dot{\hat{\bm r}} = v_0 \bm{\hat{u}} + \bm{f}(\bm{r}) + \sqrt{2D} 
    \hat{\bm r} \times \bm \xi ,
    \label{eq:motion_sphere}
\end{equation}
where $\hat{\bm r} \equiv {\bm r}/|{\bm r}|$,
while the noise $\bm \xi$ shares the same statistics as in Eq.~\eqref{eq:motion} and is interpreted in the Stratonovich sense.
Furthermore, in contrast with the Taylor-Green flow case studied in the main text, for this setup the characteristic length scale of the flow is comparable with the sphere radius: $\ell \sim R$, such that the P\'eclet number is here defined as $\Pe = R v_0/D$.

Lastly, the extension of AP and AAP to the case where motion takes place on a sphere straightforwardly follows from the presentation in the text, as it only requires us to generalize the definitions of the distance and relative direction between two points of interest.  
On a sphere of radius $R$, the shortest distance between the points $\bm{r}$ and $\bm{r}_c$---also known as great-circle distance---is defined as 
\begin{equation}
|\Delta r| \equiv R \arccos\left( \frac{\bm{r}\cdot\bm{r}_c}{R^2} \right) .
\end{equation}
The direction $\hat{\bm n}$ from $\bm{r}$ to $\bm{r}_c$ at the point $\bm{r}$ is likewise defined from the shortest arc linking the two points, 
namely 
\begin{equation}
\hat{\bm n} \equiv \frac{(\bm{r}\times\bm{r}_c)\times\bm{r}}{|(\bm{r}\times\bm{r}_c)\times\bm{r}|}.
\end{equation}
The desired heading direction $\bm{\hat{u}}$ for AP and AAP was then obtained from rotations of $\hat{\bm n}$ around the axis set by $\bm r$ using Rodrigues rotation formula \cite{Rodrigues1840}:
\begin{equation*}
    \bm{\hat{u}} = {\cal G}(\Delta r)\hat{\bm n}  \pm (\bm{r} \times \hat{\bm n})\sqrt{1- {\cal G}^2(\Delta r)} + \bm{r} (\bm{r} \cdot \hat{\bm n})(1 -  {\cal G}(\Delta r)) \ ,
\end{equation*}
where the protocol function $ {\cal G}(\Delta r)$ is defined by Eq.~\eqref{eq:policy_general} and the $\pm$ sign ensures that $\bm{\hat{u}} \cdot \hat{\bm t} \ge 0$.

\section*{Acknowledgements}
This work has received support from the Max Planck School Matter to Life and the MaxSynBio Consortium, which are jointly funded by the Federal Ministry of Education and Research (BMBF) of Germany, and the Max Planck Society.


\appendix

\begin{figure}[t]
\includegraphics[width=\linewidth]{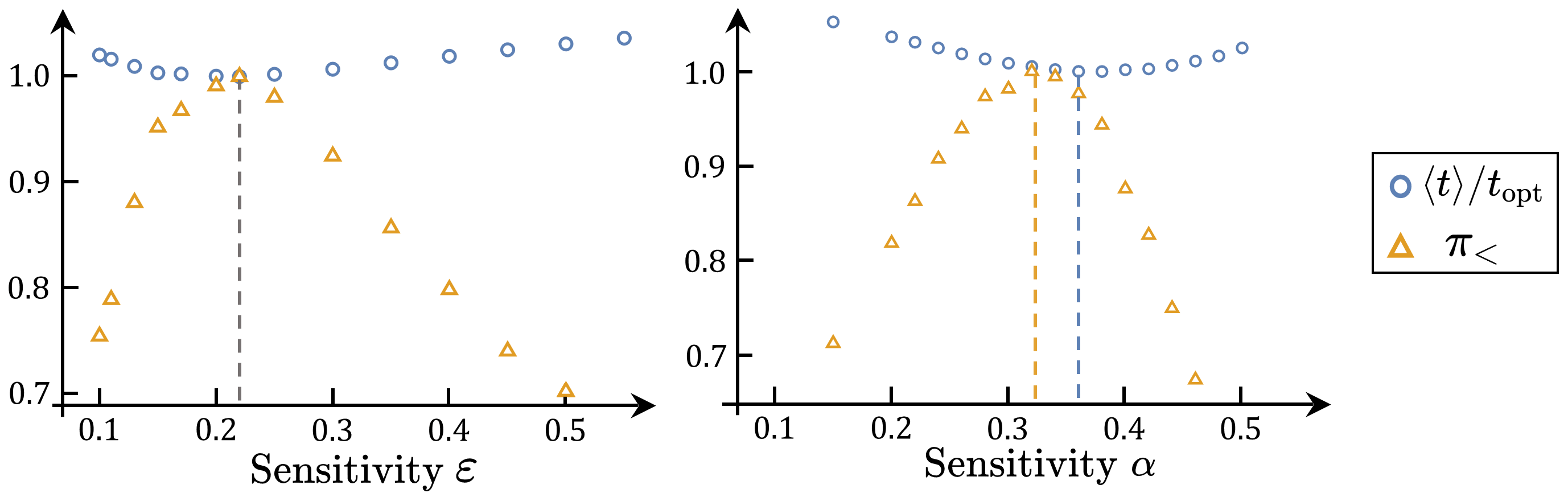}
\caption{Examples of the curves leading to the determination of the optimal sensitivities $\varepsilon$ (AAP, left) and $\alpha$ (AP, right) at $\Pe = 400$ and $\gamma=0.7$. Blue circles: mean arrival time $\langle t \rangle$ in units of the optimal time in absence of noise $t_{\rm opt}$. Orange triangles: probability of getting to the target in a time $t<t_{\rm opt}$. Here, both observables are normalized by their maximum/minimum value. In both plots the corresponding optimal sensitivity values are indicated by the vertical dashed lines.} 
\label{fig:S1}
\end{figure}

\section{An alternative policy performance indicator}
\label{sec:1}

As customary in the context of optimal navigation, the primary performance indicator is the mean arrival time at the target $\avgt$. The corresponding results shown in the main text, reveal that the new simple protocols introduced exhibit performances actually very close to that of OP regardless of the relative flow strength.
However, as our numerical simulations give us access to the full arrival time distribution we may get further insights into the performance of a navigation protocol by considering additional observables which 
also take explicitly into account the effect of fluctuations.

The specific choice we have made here stems from the following remark: owing to the presence of thermal fluctuations, the active particle may reach the target in less time than at $D=0$. The frequency of these events can be quantified by looking at the probability of arriving before the optimal time $t_{\rm opt}$ in the absence of noise (defined by the Zermelo solution, see the main text): ${\rm Prob}(t<t_{\rm opt})\equiv \pi_<$. The latter is a measure of how the policies manage to optimize the effect of fluctuations by maximizing the frequency of small arrival time events.
 Note that since in general $\avgt \ge t_{\rm opt}$, $\pi_<$ is bounded by $\tfrac{1}{2}$ from above.\\

The new navigation policies introduced in the main text both depend on a free parameter representing the protocol sensitivity. Figure~\ref{fig:S1} shows the policies performances measured from $\avgt$ and $\pi_<$ as function of sensitivity for both the adaptive aligning policy (AAP, left) and the aligning policy (AP, right). 
Similarly to the mean arrival time, 
the performance indicator $\pi_<$ allows to obtain a clear optimal sensitivity value for both AP and AAP.
Moreover, the optimal values obtained independently from $\avgt$ and $\pi_<$ generally coincide for AAP while some small differences are observed for AP. 
In the latter case, choosing either of the two estimates does not lead to significant variations of the values of $\avgt$ and $\pi_<$, and the corresponding arrival time distributions do not differ significantly.
\\

\begin{figure}[t!]
\includegraphics[width=\linewidth]{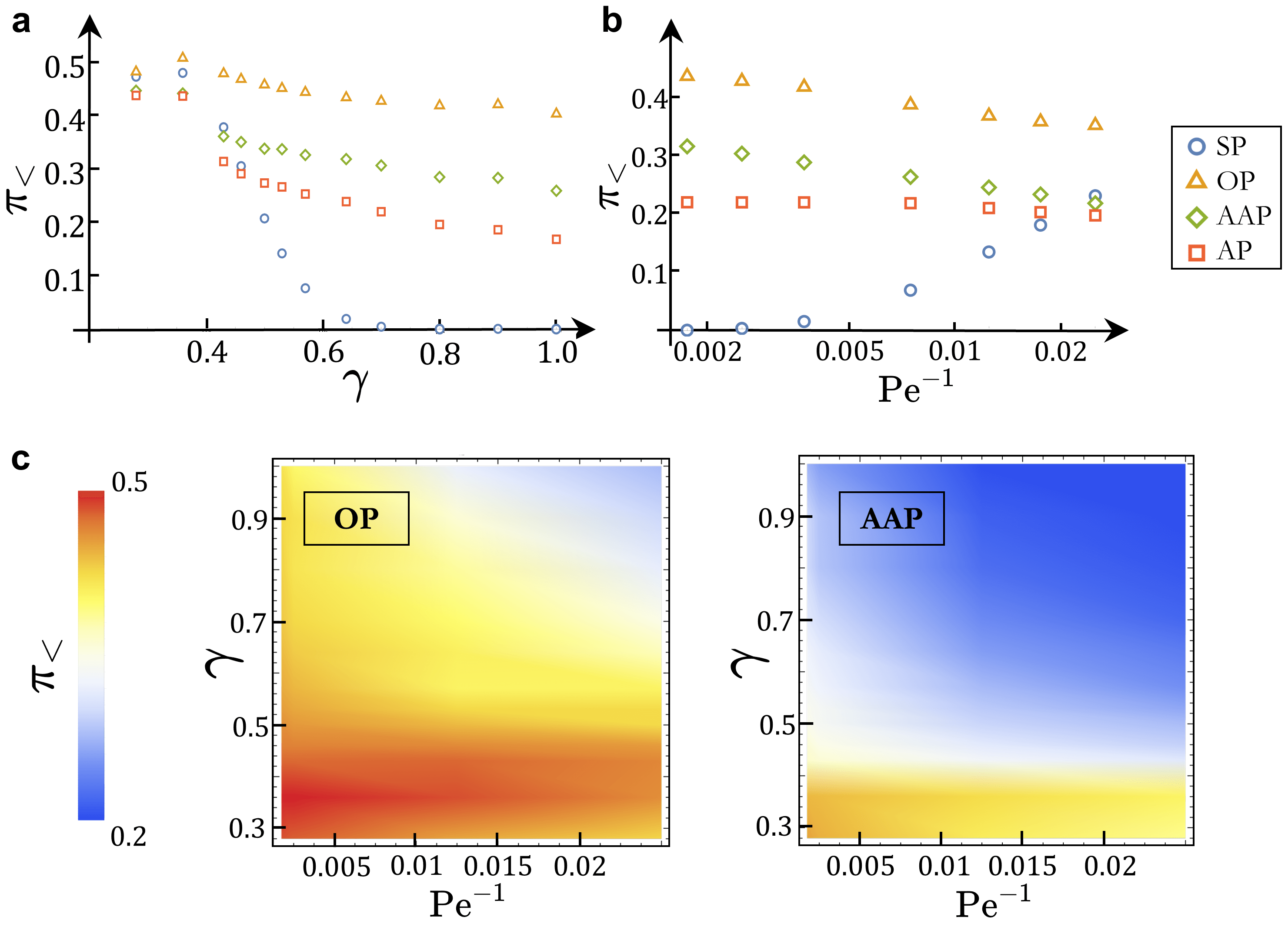}
\caption{Probability of reaching the target before the optimal time in absence of noise ($t_{\rm opt}$) as function of (a) the relative flow strength $\gamma$ at $\Pe=400$ and (b) the inverse of the P\'eclet number at $\gamma=0.7$ (horizontal axis in log scale). (c) Colour maps showing the performance of OP (left) and AAP (right) as function of both the relative flow strength $\gamma$ and the P\'eclet number $\Pe$. The colour gradient represents the probability of reaching the target faster than in absence of noise from low (blue) to high (red).
All data in (a)-(c) were averaged over $10^5$ independent trajectories.} 
\label{fig:S2}
\end{figure}

\begin{figure*}[t]
\includegraphics[width=0.85\linewidth]{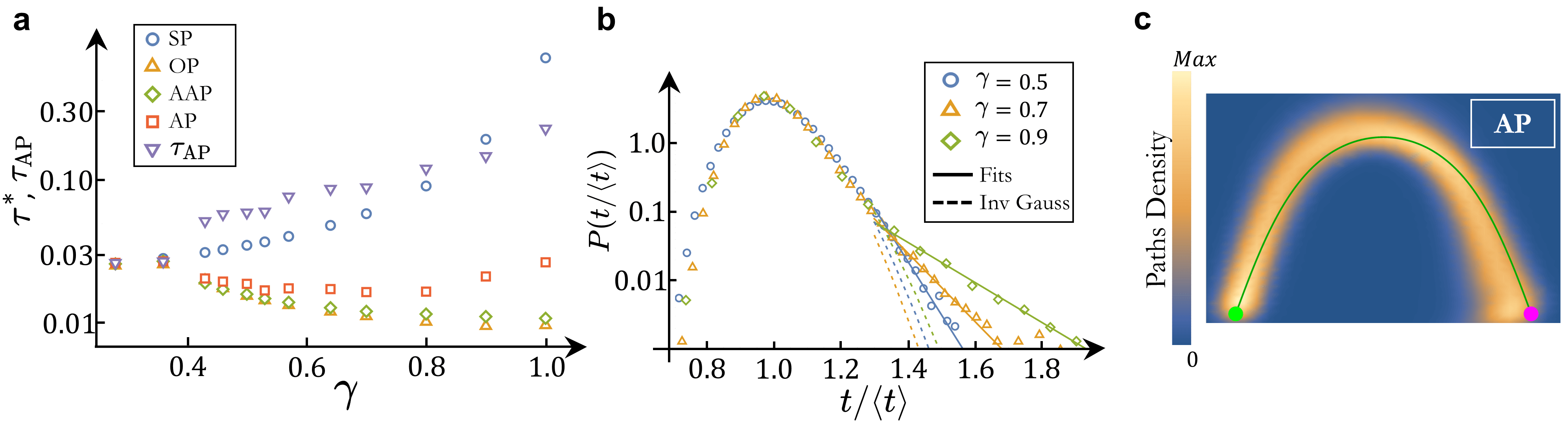}
\caption{(a) Arrival time variance to square mean ratio as function of the relative flow strength $\gamma$ at $\Pe = 400$ for the four policies: straight policy (blue circles), optimal policy (yellow triangles), adaptive aligning policy (green diamonds) and aligning policy (red squares). The inverted triangles are the values obtained from the fit of the large-time tails of AP distribution (vertical axis in log scale).
(b) Arrival time probability distributions as function of the normalized time $t/\avgt$ corresponding to AP for three different values of the relative flow strength. Solid lines show the exponential fit at large times, while the dashed lines correspond to the tails predicted by the inverse Gaussian law~\eqref{eqS:Inv_G} using the values $\avgt$ and $\sigma$ from the data. 
(c) Heat map of $10^3$ stochastic trajectories obtained from numerical simulations at $\gamma=0.7$ and $\Pe=400$ of AP.
The green curve shows the Zermelo path connecting the initial point $\bm{r}_0 = -\ell/2\hat{\bm e}_x$ (green circle) and the target $\bm{r}_T = \bm 0$ (magenta circle). 
All data in (a) and (b) are averaged over $10^5$ independent trajectories.
}
\label{fig:S3}
\end{figure*}

\noindent {\it Performance and robustness assessment.}
In Fig. \ref{fig:S2}a we show the probability $\pi_<$ as a function of the relative flow strength $\gamma$. On the one hand, the policies performances show trends analogous to those reported in the main text for the analysis of the mean arrival time, with a preserved hierarchy at strong flows. As expected, in order of increasing performance we also find here SP, AP, AAP and OP.
On the other hand, the differences between the various protocols appear more striking. For example, the probability that a swimmer following AAP reaches the target in a shorter time than $t_{\rm opt}$ is around $35\%$ lower than that of a swimmer following OP at $\gamma=1$, while the corresponding mean arrival times deviate by only a few percent (see main text).
As shown in Fig.~\ref{fig:S2}b, similar conclusions can be reached examining the behaviour of $\pi_<$ as function of the P\'eclet number.
Namely, all OP, AAP and AP show a slight decrease in performance upon increasing the strength of fluctuations, while SP becomes more advantageous at large noises.
For completeness, we also show in Fig.~\ref{fig:S2}c the heat maps comparing the values of $\pi_<$ for OP and AAP as function of $\gamma$ and $\Pe$.
Here again, the behaviour of the performance indicator is similar to that of the mean arrival time.

Overall, the new indicator $\pi_<$ is largely dominated by the mean arrival time, although it accounts explicitly for the effect of fluctuations and provides a more refined evaluation of the protocols performances. $\pi_<$ therefore leads to qualitatively analogous conclusions regarding the policies efficiency, which highlights the robustness of the results presented in the main text.

\section{Crossover from Inverse Gaussian to exponential decay for AP}
\label{sec:2}

In the main text, we showed that for most flow and noise strengths, the arrival time distributions of AAP and OP essentially follow inverse Gaussian laws, whose expression we report here for convenience:
\begin{equation}
    P(\tau) = \sqrt{\frac{\avgt^2}{2\pi\sigma^2 \tau^3}} \exp\biggr[-\frac{\avgt^2(\tau-1)^2}{2\sigma^2\tau}\biggr] \ ,
    \label{eqS:Inv_G}
\end{equation}
where $\sigma^2$ denotes the corresponding variance and $\tau=t/\avgt$. 

However, we observed that while the the arrival time distributions for AP are well described by the inverse Gaussian law at small $\gamma$, significant deviations were found when increasing the intensity of the flow. 
In particular, these deviations mainly lie  
in the large-time tail of the distribution 
that shows an exponential decay with characteristic time $\tau_{\rm AP}$ generally larger than the value $2\sigma^2/\avgt^2$ predicted by the inverse Gaussian law~\eqref{eqS:Inv_G} (see Fig.~\ref{fig:S3}b).
Defining $\tau^* \equiv 2\sigma^2/\avgt^2$ as the variance to square mean ratio of arrival time 
for each protocol, 
its scaling with the flow strength is shown in Fig.~\ref{fig:S3}a for all policies.
As a sign that AAP and OP on average lead the swimmer to travel faster as the flow strength is increased, $\tau^*$ decays with $\gamma$ for these two policies.
In contrast, swimmers following SP always travel counter-flow and are thus get slower on average as $\gamma$ increases. 
They are thus more subject to fluctuations, such that for SP $\tau^*$ grows with $\gamma$.
Lastly, for AP $\tau^*$ undergoes a crossover 
from a decay with flow strength at small $\gamma$, 
to a growth with $\gamma$ at large flows (red squares in Fig.~\ref{fig:S3}a).
On the other hand, the value $\tau_{\rm AP}$ obtained from the large-time tails of the distribution always grows with $\gamma$, similarly to SP (purple inverted triangles in Fig.~\ref{fig:S3}a).
These observations can be rationalized from the heat map of AP trajectories shown in Fig.~\ref{fig:S3}c. 
Namely, it shows that most of the trajectories end at the left of the target, such that the swimmers have to travel counter-flow 
and are thus generally slower in the final stretch. 
This suggests that, for AP, the events characterized by $\tau>1$ are dominated by this last part of the swimmer's journey, ultimately caused by the overall pronounced asymmetry with respect to the Zermelo path.

\begin{figure*}[t]
\includegraphics[width=0.95\linewidth]{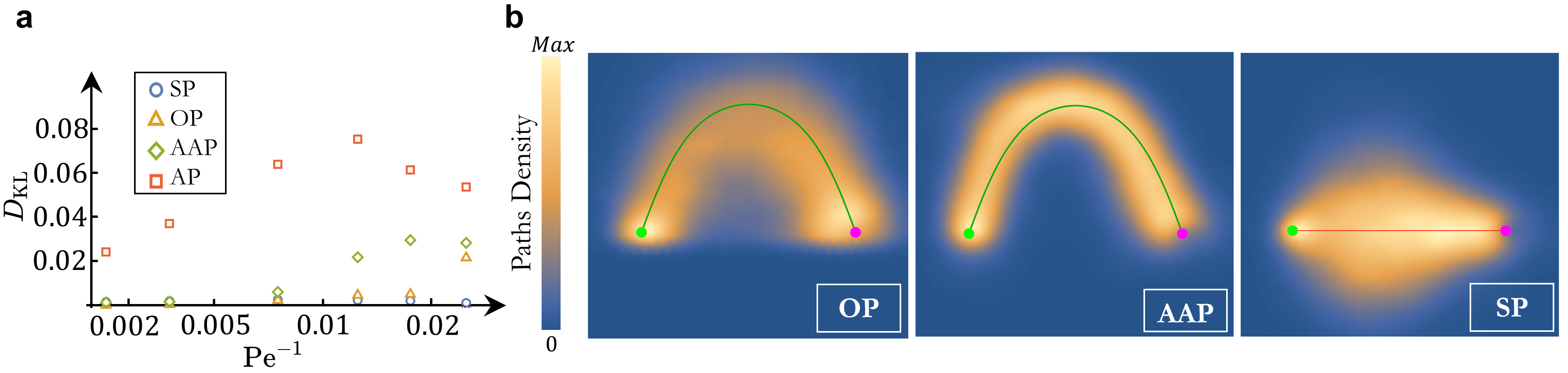}
\caption{(a) Kullback-Leibler divergence $D_{\rm KL}$ between the numerical and theoretical arrival time distributions as function of the inverse of the P\'eclet number at $\gamma=0.7$ (horizontal axis in log scale). (b) Heat maps of $10^3$ stochastic trajectories obtained from numerical simulations of OP (left), AAP (center) and SP (right) at $\Pe = 40$ and $\gamma=0.7$. The solid green lines here represent the $D = 0$ Zermelo path connecting the initial point $\bm{r}_0 = -\ell/2 \hat{\bm e}_x$ and the target $\bm{r}_T = \bm 0$, marked with a green and magenta circle, respectively. The solid red line is instead the straight path linking the two points.} 
\label{fig:S4}
\end{figure*}

\section{Large fluctuations and the inverse Gaussian law}
\label{sec:3}

The correspondence observed between the inverse Gaussian law and the arrival time distributions of OP, AAP and SP is due to the fact that, as long as the thermal fluctuations are reasonably small, these policies guide the particle along a fictional path, thus reducing the system to an active Brownian particle navigating in a quasi-1D environment.

However, this analogy may break down in the regime of strong fluctuations. This can be quantified as in the main text computing the Kullback-Leibler divergence $D_{\rm KL}$ between the distributions obtained from the direct numerical simulations $P_{\rm num}$ and those predicted by the inverse Gaussian law \eqref{eqS:Inv_G} with parameters $\avgt$ and $\sigma$ extracted from the data. More specifically, given a discretization ${t_i}$ of the relevant time interval, the Kullback-Leibler divergence has been calculated as $D_{\rm KL} = \sum_i P_{\rm num}(t_i) \ln[P_{\rm num}(t_i)/P(t_i)]$. Figure~\ref{fig:S4}a shows the corresponding results obtained for all the policies. Remarkably, the arrival time distributions of both OP and AAP turn out to be inverse Gaussian for a wide range of noise amplitudes, with significant deviations arising only from $\Pe^{-1} \gtrsim 0.01$. As shown in the heat maps in Fig.~\ref{fig:S4}b (left and central panels), when fluctuations are strong the stochastic trajectories are in fact less focused around the optimal path and also more asymmetrically distributed around it. As already discussed in the previous section, this asymmetry leads to larger tails in the probability distributions (data not shown) and therefore to the observed deviations from the inverse Gaussian law.

In this scenario, the shape of SP distribution turns out to be the most robust to fluctuations. This can be better understood by looking at the corresponding heat map in Fig.~\ref{fig:S4}b (rightmost panel). Despite being more dispersed, the stochastic trajectories still look symmetrically distributed around the straight path. This is strictly related to the flow symmetry with respect to the line connecting the starting point with the target (to this end, please refer to Fig. 1a in the main text).

\section*{References}
\bibliographystyle{apsrev4-2}
\bibliography{bib_nat_comm}

\end{document}